\documentstyle[preprint,aps,epsfig,amstex]{revtex}  
 
\tighten       
 
\begin{document} 

\title{Geometry effects at conductance quantization in quantum wires}
\author{G. Kiesslich, A. Wacker, and E. Sch{\"o}ll}
\address{Institut f{\"u}r Theoretische Physik, Technische Universit{\"a}t
Berlin, Hardenbergstra{\ss}e~36, D-10623 Berlin, Germany}

\date{physica status solidi, sceduled for Vol. 216/2, page R5 (1999))\\
(http://www.wiley-vch.de/berlin/journals/pss/rapid/contents/index.html\#99-061}

\maketitle

In Refs.\cite{YAC96,ROT99b}  the fabrication of quasi-onedimensional
electronic systems (quantum wires) by {\em cleaved edge overgrowth}
(CEO) in combination with a gate potential has been reported.
Measured mean free
paths \cite{YAC96} of about 10$\mu$m indicate that
the electrons pass the wire
ballistically. Therefore the Landauer formula suggests the
linear-response conductance $G=\frac{2e^2}{h}M$, where $M$ is the number
of transverse modes in the wire which can be reduced by increasing the gate
 potential $\vert V_{g}\vert$. In contrast, measurements
\cite{YAC96,ROT99b} show
plateaus below the theoretical value. Possible explanations may be based
on either many particle interactions in the wire or geometry
effects causing scattering at
the ends of the wire.
Our goal is to analyze the magnitude of those geometry effects neglecting
interactions. Specifically, we investigate the
 influence of geometry and spatial potential landscape
on the interference between wire states and contact states.

Our calculations were performed applying the method of equilibrium
Green's functions
following Ref.~\cite{DAT+FER}. Within the discretization in tight binding
approximation ($a=7.5$ nm) of the Hamiltonian and Dirichlet boundary
conditions the Green's functions are finite matrices. The leads
(source and drain)
are treated in terms of self-energy.
We use the effective mass $0.067m_e$.

\begin{figure}
\noindent\epsfig{file=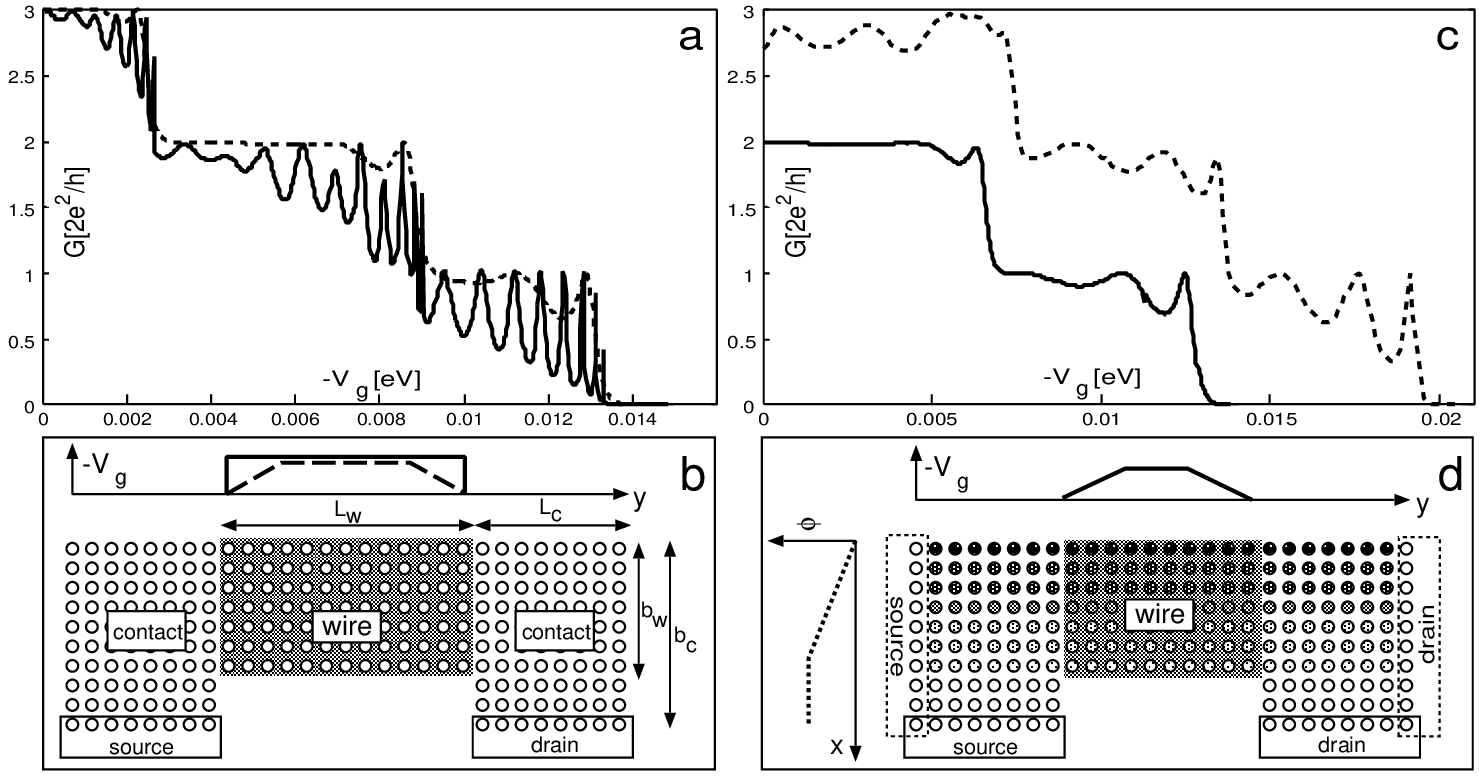,width=\textwidth}
\caption{Conductance $G$ versus gate voltage $V_g$ and scheme of
CEO devices with different potential landscapes. (a,b)
Rectangular (solid line) and trapezoidal (dashed line)
gate potential in the wire (shaded region) for
$L_c=300$ nm, $L_w=300$ nm.
(c,d) Additional attracting potential $\phi$ close to the CEO interface
for different lead configurations: U-shaped (solid source/drain box
in (d) with $b_c=53$ nm, solid  curve in (c)) and linear
(dashed box in (d) with $b_c=150$ nm, dashed  curve in (c)) 
for $L_c=300$ nm, $L_w=150$ nm.
The lengths are not drawn to scale in (b,d).}
\label{3mode}
\end{figure}

Fig.~\ref{3mode} shows the results of our calculations, where
$V_g$ is applied over the whole range of the wire of length $L_w$ 
and width $b_w=53$ nm. In Fig.~\ref{3mode}a
(geometry as Fig.~\ref{3mode}b)
three plateaus appear because there are three transversal
modes in the wire below the Fermi energy of 15 meV for $V_g=0$.
Here, the width of the contacts, $b_c$,
does not affect the transmission for $b_c\ge b_w$, thus we choose
$b_c = b_w$. The solid line in Fig.~\ref{3mode}a correspond to a
rectangular potential step
between contacts and wire. The oscillations in the conductance
plateaus are well known as quantum mechanical transmission through a
barrier and their frequency scales with the length of the wire.
The dashed line in Fig.~\ref{3mode}a follows from a calculation
for a trapezoidal gate potential with
a linear ramp over 30 nm at both ends of the wire.
The oscillations almost disappear and almost complete quantization in
every plateau is found. Thus  we are still in the adiabatic
regime \cite{GLA88} and no reflections
at the boundary between contact and wire appear.
However, in the solid curve the averaged
plateau height of the lowest mode is
clearly below the universal value in agreement with
recent measurements in V-groove wires \cite{KAU99}. If the
potential changes abruptly in the transition between
contacts and wire the mismatch between different states
 increases and there is a reduction in transmission for lower modes, which
essentially affects the first jump at high $V_g$ while the subsequent jumps
become closer to $2e^2/h$.

Furthermore we considered  an additional
attracting potential $\phi$ (groove) close to the CEO interface 
(top boundary of Fig.~\ref{3mode}d)  which extends over the whole
length $L_w+2L_c$ and drops linearly  
along the x direction by 10 meV on the scale of 53 nm.
The solid line in Fig.~\ref{3mode}c refers to
the U-shaped lead configuration 
(solid boxes for source and drain in Fig.~\ref{3mode}d). 
In this case only two plateaus are observed: the lowest
mode in the groove has an energy below the
contact potential and is not accessible from the leads. 
If we change the configuration of leads (dashed boxes in
Fig.~\ref{3mode}d), three plateaus are observed again,
see  the dashed curve in Fig.~\ref{3mode}c. 
Alternatively, one can obtain the same effect 
by placing some randomly distributed localized potentials in the groove 
regions, which act as scatterers.
Thus, it is essential from which direction the electrons are injected 
into the device if this additional groove potential $\phi$ is present.
 
We have studied the geometry effects in cleaved
edge overgrowth  structures on the transmission through quantum wires.
As long as the energy of the modes in the wire
is {\em above contact level}, the junction between a wide
contact and narrow wire has almost no influence on the transmission
while a sharp potential step between contact and wire
strongly affects the conduction. In contrast, modes with energies
{\em below the contact level} may be difficult to access. Here the position
of the external leads as well as the presence of scattering is of crucial
importance.

\end{document}